\documentclass[reprint,pra,aps,amsmath,amssymb, twocolumn,superscriptaddress, raggedbottom]{revtex4-2}

\usepackage{graphicx}
\usepackage{dcolumn}
\usepackage{bm}
\usepackage{bbm}
\usepackage[colorlinks=true, urlcolor=blue,citecolor=blue,anchorcolor=blue,breaklinks=true]{hyperref}
\usepackage[capitalise]{cleveref}
\usepackage{dsfont}
\usepackage{mathtools}
\usepackage{amsthm, thmtools}
\usepackage{amsmath}
\usepackage{empheq}
\usepackage[ruled]{algorithm2e}
\usepackage{xcolor}
\usepackage[normalem]{ulem}

\newcommand{\microspace}{\mspace{0.5mu}}

\def\<{\langle}
\def\>{\rangle}
\def \lket {\left|}
\def \rket {\right\rangle}
\def \lbra {\left\langle}
\def \rbra {\right|}
\newcommand{\ket}[1]{\lket\microspace #1 \microspace\rket}

\newcommand{\bra}[1]{\lbra\microspace #1 \microspace\rbra}

\newcommand{\comm}[2]{\left[#1, #2\right]}
\newcommand{\der}[2]{\frac{d\, #1}{d\, #2}}

\newcommand{\fder}[2]{\frac{\delta\, #1}{\delta\, #2}}
\newcommand{\avg}[1]{\left\langle#1\right\rangle}
\usepackage{mathtools}

\usepackage{bm}

\begin{document}
\title{FOCQS: Feedback Optimally Controlled Quantum States}

\author{Lucas~T.~Brady}
\email{Lucas.T.Brady@nasa.gov}
\affiliation{Quantum Artificial Intelligence Laboratory, NASA Ames Research Center, Moffett Field, California 94035, USA}

\author{Stuart Hadfield}
\affiliation{Quantum Artificial Intelligence Laboratory, NASA Ames Research Center, Moffett Field, California 94035, USA}
\affiliation{USRA Research Institute for Advanced Computer Science (RIACS), Mountain View, CA 94043, USA}

\date{\today}
\begin{abstract}
Quantum optimization, both for classical and quantum functions, is one of the most well-studied applications of quantum computing, but recent trends have relied on hybrid methods that push much of the fine-tuning off onto costly classical algorithms.  Feedback-based quantum algorithms, such as FALQON, avoid these fine-tuning problems but at the cost of additional circuit depth and a lack of convergence guarantees.  In this work, we take the local greedy information collected by Lyapunov feedback control and develop an analytic framework to use it to perturbatively update previous control layers, similar to the global optimal control achievable using Pontryagin optimal control.  This perturbative methodology, which we call Feedback Optimally Controlled Quantum States (FOCQS), can be used to improve the results of feedback-based algorithms, like FALQON.  Furthermore, this perturbative method can be used to push smooth annealing-like control protocol closer to the control optimum, even providing and iterative approach, albeit with diminishing returns.  In numerical testing, we show improvements in convergence and required depth due to these methods over existing quantum feedback control methods.
\end{abstract}

\maketitle

\section{Introduction}

Quantum Optimization has long held a place as one of the more popular applications, theoretically and experimentally, for near-term quantum computers.  Despite this popularity, quantum optimization suffers from severe limitations, often requiring difficult and lengthy calculations and procedures to achieve any noticeable or relevant quantum advantage.  The current state of the field largely ignores these computationally taxing steps because they are still relatively small and feasible for near-term algorithms.

This prevalence of optimization problems in quantum computing dates back at least to quantum adiabatic computing \cite{Kadowaki1998, Farhi2000} which has performance guarantees based on the quantum adiabatic theorem \cite{Jansen2007, Lidar2009}.  Adiabatic Computing is universal for quantum computing \cite{Aharonov2005}, but the corresponding proof requires arbitrary operations whereas experimental adiabatic quantum computers often rely on more limited stoquastic operations whose effectiveness is more dubious \cite{Brady2016,Jiang2017,Bringewatt2020,Crosson2021}.  Even when quantum advantage is possible with adiabatic quantum computation, it often requires a priori or numeric optimization of the annealing schedule to realize that advantage \cite{Roland2002,Rezakhani2009,Jarret2019}.

These limitations in Adiabatic Quantum Computing and its generalization, Quantum Annealing, are partly mitigated by the Quantum Approximate Optimization Algorithm (QAOA) \cite{Farhi2014,Hadfield2019} and the Variational Quantum Eigensolver (VQE) \cite{Peruzzo2014}.  Both these algorithms approach quantum optimization by creating a parameterized quantum circuit (based off Quantum Annealing for QAOA and based off chemical ans\"atze for VQE) and then running an outer loop of classical optimization to set those parameters so that the quantum protocol minimizes some cost function.  These algorithms, as well as variational forms of quantum annealing \cite{Roland2002,Rezakhani2009,Matsuura2020,Wurtz2022,Unsal2022} and more general variational quantum algorithms \cite{Brady2021,Brady2021b}, achieve better performance than more naive implementations of quantum annealing but at the cost of this extra layer of classical optimization.  This classical outer loop optimization can itself be NP-Hard \cite{Bittel2021}, so while near term applications need few enough parameters to be feasible, variational quantum algorithms will not be feasible for quantum optimization on larger quantum problems.

So then one of the biggest questions facing quantum optimization is how to achieve a portion of the advantages and speedups inherent in variational quantum algorithms without relying on variational ans\"atze that require infeasible scaling in their number of parameters.  Focusing down on Quantum Annealing and QAOA-like procedures, there are remarkably few variants that attempt to reduce down the variational load, particularly while guaranteeing performance improvement. There exist numerous proposed variants of these algorithms, but most of them are focused on improving pre/post-processing or enhancing the variational ansatz \cite{Hadfield2019,zhu2022adaptive}.  One of the few variants to actually attempt to remove the variational parameters is FALQON \cite{Magann2022,Magann2022b} which relies on Lyapunov control, a type of feedback stabilization control to try to ``stabilize'' a quantum state into the ground state of a quantum system.

This FALQON algorithm is remarkable in that it works, often faster than unoptimized quantum annealing, and achieves fairly good results.  However, the algorithm has several primary limitations: a) it lacks guarantees that it will converge to the ground state (and often doesn't), as originally designed b) it 
gets effectively stuck 
at suboptimal solutions of the cost function and c) it 
can require orders of magnitude larger circuit depth than an equivalently performing optimized QAOA circuit would need, among others.  The first problem is a similar concern theoretically in QAOA and other variational quantum algorithms, but in practical terms, FALQON suffers from false minima more than other variational quantum algorithms.  Part of the lack of theoretical guarantee here is because Lyapunov control is usually concerned with stabilization rather than full state preparation, meaning most existing proofs of convergence assume more than we can provide in a usual quantum computing setting.
Another recent paper \cite{malla2024feedbackbased} provided an alternate formulation which included an additional catalyst Hamiltonian in the FALQON ansatz, trying to mimic counterdiabatic driving and speed up the convergence rate.

The main focus of this paper will be to improve the speed and circuit depth of feedback-based quantum algorithms, such as but not limited to FALQON and direct variants. 
This improvement comes at the cost of additional classical processing time proportional to the length of the circuit in the worst setting, but this processing can be reduced to a constant overhead in most realistic situations.  In classical control problems, Lyapunov control would be used to continuously steer a system through feedback, but the quantum setting requires re-preperation of the state after every measurement.  This setting gives us an opportunity to not just modify the most recent elements of the circuit based on new information but also update previous parts of the circuit based on new measurements.  Thus, we can leverage the destructive nature of quantum measurements to further improve the circuit and create better quantum protocols.  In a sense, Lyapunov control is akin to a greedy search, based off only information local at the end of the procedure; whereas, Pontryagin optimal control \cite{Pontryagin} is more akin to greedy search using global information from the entire history of the procedure.  Our goal will be to use the information that FALQON or similar algorithms are already collecting and use that to go back and perturbatively estimate the information we need for more global optimal control.

To that end we introduce a new quantum algorithm, termed Feedback Optimally Controlled Quantum States (FOCQS).  This algorithm uses the same information as FALQON but in a perturbative way that continues to update previous steps as the protocol grows.  Even more excitingly, this FOCQS protocol can be applied as a variational-parameter-free iteration method that can improve existing annealing-like (or Trotterized annealing-like) quantum circuits, improving them with each iteration.  This iteration method could be used on a FALQON circuit, a naive quantum annealing schedule, or the output of a previous iteration of FOCQS (although, benefits do diminish with subsequent iterations).

In the next section, Sec.~\ref{sec:FALQON}, we review the form of FALQON and discuss feedback algorithms in general.  We follow this up in Section~\ref{sec:optimal_control} by discussing other forms of optimal control theory and how they compare to the forms used by the Lyapunov control supporting FALQON.  Then in Subsection~\ref{ssec:theory}, we combine these notions of control theory and derive a perturbative method for using the information naturally collected during FALQON to further improve the procedure.  We introduce and analyze our new FOCQS algorithm and several variants in Sec.~\ref{sec:algorithms} and then present numeric results in Sec.~\ref{sec:numerics}.  Finally, we conclude in Sec.~\ref{sec:conclusion}.

\begin{algorithm}
\caption{FOCQS meta-algorithm}
\label{alg:focqs0}
\SetKwInOut{Input}{input}
\Input{$\hat{C}$ - Target Hamiltonian to optimize}
\Input{$\hat{H}(u;\hat{C})$ - Parameterized quantum Hamiltonian}
\Input{$p$ - Algorithm depth (number of layers)}
\Input{$\ket{\varphi_0}$ - Easily preparable initial state}
$u_0\gets0$\;
\For{$j\gets 0\ldots p$}{
    1) Construct unitary $\hat{U}_j = \prod_{j=0}^p e^{-i \Delta t \hat{H}(u_j)}$\;
    
    2) Prepare $\ket{\psi_j} = \hat{U}_j\ket{\varphi_0}$ \;
    
    3) Measure specified observables of $\ket{\psi_j}$\;
    
    4) Use observable data to set $u_{j+1}$\;
    
    5) Update $u_k$ for $k = 0,\dots,j$ using observables\;
}
\end{algorithm}
In Algorithm \ref{alg:focqs0}, we provide a very broad overview of our proposed methods which serve as a generalization of FALQON.
Steps (1) to (3) involve repeatedly preparing and measuring the current quantum circuit ansatz to estimate the relevant observable expectation values, and this data is then used to update the ansatz parameters in Steps (4) and (5).
The FALQON algorithm corresponds to a single observable used to update the current layer, $u_j$ which is a scalar, not a vector of parameters.  Furthermore, FALQON excludes Step (5) which is the main innovation of this paper.  Our iterative version of the algorithm replaces Step (4) with an a priori determined set of parameters.
We provide details of our procedure including more detailed algorithms and the analytics behind our update rules in the rest of the paper.

\section{FALQON} 
\label{sec:FALQON}

In this section we look at the basics of FALQON, as it was originally proposed by Magann \emph{et al.}~\cite{Magann2022}. 

In the original FALQON, we have a discretized control Hamiltonian broken into time slices, indexed by $j$.  The Hamiltonian in a time slice is given by
\begin{equation}
    \hat{H}_j = u_j \hat{B} + \hat{C},
\end{equation}
where $u_j\in[0,\infty)$.  This setting is similar to some formulations of quantum annealing and QAOA, where $\hat{C}$ is a problem Hamiltonian (often diagonal) that encodes the solution to the computational problem in its ground state.  $\hat{B}$ is a mixer Hamiltonian that is relatively simple and has an easy to prepare ground state; often $\hat{B}$ is a transverse field applied to the qubits, though our analysis applies more generally. 

Our evolution starts from the ground state of $\hat{B}$, $\ket{\varphi_0}$, and will be governed by a $p$ depth unitary ansatz:
\begin{equation}
    \label{eq:state}
    \ket{\psi_p} = \left[\prod_{j=1}^p e^{-i \Delta t \hat{H}_j}\right]\ket{\varphi_0}.
\end{equation}
For our modifications to FALQON $\Delta t$ should be a small-ish parameter since we plan to use it for perturbative Taylor expansions later on. Further Trotter approximations such as $e^{-i \Delta t \hat{H}_j}\simeq e^{-i \Delta t \hat{C}}e^{-i \Delta t  u_j \hat{B}}$ can be used as needed in numerics and experiments.

The goal of a feedback algorithm would be to take an existing $p$ layer algorithm and figure out what to attach as the next layer.  The goal of such an algorithm is to minimize $\avg{\hat{C}}$ for the state at the end of the procedure, so FALQON chooses the next layer in a way that guarantees a monotonic decrease of the value of $\avg{\hat{C}}$ while avoiding the cost of parameter search. 

To this end, we can calculate a time derivative, switching between the indices and time via $t_j = j \Delta t$.  Remember that this system will obey the Schr\"odinger equation so that
\begin{equation}
    i\der{}{t}\ket{\psi_j} = \hat{H}_j \ket{\psi_j}.
\end{equation}
Therefore, the time derivative of the energy of the current state is
\begin{align} \label{eq:phi_j0}
    \left.\der{\bra{\psi(t)}\hat{C}\ket{\psi(t)}}{t}\right|_{t=t_j}
    =  u_j\bra{\psi_j}i\comm{\hat{C}}{\hat{B}}\ket{\psi_j} \eqcolon u_j \frac{\phi_j}{\Delta t},
\end{align}
It will be useful to also express $\phi_j$ in the Heisenberg picture:
\begin{equation}
    \label{eq:phi_j}
    \phi_j = \phi(t_j) = i\Delta t\bra{\varphi_0}\comm{\hat{C}(t_j)}{\hat{B}(t_j)}\ket{\varphi_0}.
\end{equation}

The methodology of FALQON would be to use this expression to try to make the change in the energy between circuit layers strictly non-positive.  In the original formulation, Magann \emph{et al.} \cite{Magann2022} just choose to set the mixing control parameter $u_{j+1} = -\phi_j$; although, it is easy to see it suffices to set $u_{j+1} = -\text{sign}(\phi_j) |f(\phi_j)|$ for some arbitrary function $f(x)$.  Although it is reasonable to consider a bang-bang approach, coming from QAOA, in practice a bang-bang approach leads quickly to non-optimal local minima.
In practice this dependence on mostly the sign to determine the success of the algorithm lessens the sampling requirements when implementing these kind of algorithms on real hardware.  Greater precision in measurements is helpful for determining the strength of the proposed control field, but ultimately, sampling procedures do not need to aim for much more accuracy than just the sign of the proposed control parameter.

We instead choose to follow FALQON and  set
\begin{equation}
    \label{eq:FALQON_con_0}
    u_{j+1} = 
        \begin{cases}
            0 & \phi_j>0\\
            -\phi_j & \phi_j\leq0\\
        \end{cases}.
\end{equation}
This choice assumes that the control field cannot go negative due to limitations in the hardware.  If there are no such limitations on the sign of the control field, it is simple to set 
$u_{j+1}=-\phi_j$.  We choose a nonnegative control field in our numerics, but none of the analytics or methodolgies rely on this choice.

Finally, we reemphasize that it is easy to see from Eq.~\ref{eq:phi_j0} that if the quantum algorithm has converged to any eigenstate of $C$, then $\phi_j$ will be zero identically. Hence, the above methods based on the derivative alone will have difficulty escaping local minima, which motivates our generalized algorithms to follow.

In Appendix \ref{app:bounded}, we discuss a novel modification to FALQON to use bounded control fields, similar to QAOA and other forms of quantum annealing.  This method is less useful for our later perturbative approaches, but we present it for completeness.

\section{Optimal Control}
\label{sec:optimal_control}

We next consider improving the form of FALQON listed above using ideas from optimal control theory. In particular, we propose to use the measurement data obtained throughout the overall procedure to update all circuit parameters not just the current layer.  Despite focusing on just FALQON variants, this methodology would be applicable to other forms of quantum feedback based control.  

For ease of notation, we define the function 
\begin{equation}
    J(t) \coloneq \avg{\hat{C}}_t
\end{equation}
as the expectation value of the cost function at a given time, $t$, during our procedure.  Alternative versions of these algorithms could be developed using different objective functions, but this objective, $J(t)$, aligns with standard approaches to quantum optimization.

From optimal control \cite{Pontryagin,Brady2021}, a quantity of importance for finding the optimal values of the control function is the functional derivative of the objective function at the final time $t_f$ with respect to the control function at time $t$:
\begin{equation}
    \Phi(t,t_f) \equiv \fder{J(t_f)}{u(t)}.
\end{equation}

In Ref.~\cite{Brady2021}, this quantity was calculated using a Lagrange Multiplier, which is just a fictitious state evolving backwards in time, but for our purposes, this quantity is more easily represented in the Heisenberg picture.  For the unbounded control function, the result would be
\begin{equation}
    \label{eq:Phij}
    \Phi_j(t_f) \coloneq \Phi(t_j,t_f) = i\Delta t\bra{\varphi_0}\comm{\hat{C}(t_f)}{\hat{B}(t_j)}\ket{\varphi_0}.
\end{equation}

The intriguing thing to note here is that $\Phi(t_j,t_j) = \phi(t_j)$ from Eq.~(\ref{eq:phi_j}).  This $\Phi(t_j,t_f)$ can be used as the gradient of the cost function with respect to the control parameter $u_j$.  We would like to run a feedback based algorithm that can use the information it has collected to not only add a final step but also update previous steps.  This gradient would be one possible way of doing this.

The difficulty with $\Phi(t_j,t_f)$ is that it is hard to calculate.  There are ways to measure this, either using ancilla qubits \cite{Li2017} or parameter-shift rules \cite{LiJ2017,Mitarai2018}.  Both of these require additional resources and measurements, and the parameter-shift rule especially is not very practical when dealing with global rotations like $\hat{B}$ and $\hat{C}$ (they are designed for parameters multiplying products of Pauli operators, not generally sums of Pauli products).

\subsection{Derivative Approximations}
\label{ssec:theory}

Instead, we are going to use the fact that we have access to the full history of $\phi(t_j)$ to make estimates of what $\Phi(t_j,t_f)$ should be.  The algorithmic goal would be to provide a protocol for not just adding new layers onto the unitary ansatz but updating previous layers.

One caveat before we start: if we are going to be updating previous layers, that is going to change the values of $\phi(t_j)$ away from what we have previously measured.  This could be problematic in practice, but for the remainder of this section, we assume that we have perfect access to these quantities.  In a future section, we will discuss various ways of mitigating this degradation in our information.  All of the work in this section will be perturbative anyway, so none of this will hold too far into the past regardless.

To begin our perturbative analysis, we can look at the Taylor expansion of $\Phi$ about its second argument at time $t_j$:
\begin{align}
    \Phi(t_j,t_f) =& \Phi(t_j,t_j) + (t_f-t_j)\left.\der{}{x}\Phi(t_j,x)\right|_{x=t_j} \\\nonumber&+\mathcal{O}((t_f-t_j)^2\Delta t).
\end{align}
The extra $\Delta t$ dependence in the Big-O notation is because $\Phi$ itself is proportional to $\Delta t$.

Let's note also that we can separate out the different time dependencies here by remembering that
\begin{equation}
    \Phi(t,t_f) = \fder{J(t_f)}{u(t)}~~\&~~\phi(t) = \fder{J(t)}{u(t)}.
\end{equation}
This was not how we derived $\phi(t)$, but it is a valid way of representing the resulting quantity.  Using these rewritings, we can express our perturbation theory result as
\begin{align}
    \Phi(t_j,t_f) = &\phi(t_j) + (t_f-t_j)\left.\der{}{x}\fder{J(x)}{u(t_j)}\right|_{x=t_j} 
    \\\nonumber&+\mathcal{O}((t_f-t_j)^2\Delta t).
\end{align}

At this point we are going to make a few more approximations and assumptions, some of them on shakier ground.
 
First, let's approximate the derivative here via a two-point forward difference quotient:
\begin{equation}
    \left.\der{}{x}\fder{J(x)}{u(t_j)}\right|_{x=t_j} = \frac{1}{\Delta t}\left(\fder{J(t_{j+1})}{u(t_j)}-\fder{J(t_{j})}{u(t_j)}\right)+\mathcal{O}(\Delta t^2).
\end{equation}
The second quantity here is just our known $\phi_j$, but the first one needs some more attention.  
We will define
\begin{equation}
    \tilde{\phi}(t_j) \equiv \fder{J(t_{j+1})}{u(t_j)},
\end{equation}
which then means that
\begin{align}
    \label{eq:Phi_approx}
    \Phi(t_j,t_f) = &\phi(t_j) + \frac{t_f-t_j}{\Delta t}(\tilde{\phi}(t_j)-\phi(t_j)) 
    \\\nonumber&+\mathcal{O}\left((t_f-t_j)^2\Delta t,\Delta t^3\right).
\end{align}
Again an extra factor of $\Delta t$ showed up in the Big-O notation thanks to the dependence of $\Phi$ and $\phi$ on this quantity.

This $\tilde{\phi}$ quantity is now of interest to us.  It is a trickier quantity to calculate and estimate since it is asking what effect the control function at one time step has on the objective function evaluated at the next time step.  The Schr\"odinger equation should give a continuous and differentiable result for $J(t)$, but the same cannot be said for $u(t)$ which is allowed to be piece-wise continuous.  Therefore, we cannot use nice approximations of $u(t)$ via its derivatives to make our lives easier here.

What we can do is go back to the form of $J(t)$ itself and do this functional differentiation manually.  We can express $J(t_{j+1})$ as
\begin{align}
    J(t_{j+1}) &= J(t_j + \Delta t) 
    \nonumber\\&= \bra{\psi_j}e^{i\Delta t(u_{j+1}\hat{B}+\hat{C})}\hat{C}e^{-i\Delta t(u_{j+1}\hat{B}+\hat{C})}\ket{\psi_j}.
\end{align}
This form hides the dependence on all previous $u_i$ in $\ket{\psi_j}$, but all the $u_j$ dependence sits right on the surface layer of $\ket{\psi_j}$.  This makes the differentiation easy
\begin{align}
    &\tilde{\phi}(t_j) = \fder{J(t_{j+1})}{u(t_j)}
    \nonumber\\
        & =i\Delta t\bra{\psi_j}
            \comm{\hat{B}}
                {e^{i\Delta t(u_{j+1}\hat{B}+\hat{C})}\hat{C}e^{-i\Delta t(u_{j+1}\hat{B}+\hat{C})}}
        \ket{\psi_j}.
\end{align}

Next, we are going to make a small angle approximation with these exponentials, based off the first order Baker-Campbell-Hausdorff equation, so that
\begin{align}
    &\comm{\hat{B}}
                {e^{i\Delta t(u_{j+1}\hat{B}+\hat{C})}\hat{C}e^{-i\Delta t(u_{j+1}\hat{B}+\hat{C})}}\nonumber\\
    =&\comm{\hat{B}}
                {e^{i\Delta t u_{j+1}\hat{B}}e^{i\Delta t\hat{C}}\hat{C}e^{-i\Delta t\hat{C}}e^{-i\Delta t u_{j+1}\hat{B}}}+\mathcal{O}(\Delta t^2)\nonumber\\
    =&\comm{\hat{B}}
                {e^{i\Delta t u_{j+1}\hat{B}}\hat{C}e^{-i\Delta t u_{j+1}\hat{B}}}+\mathcal{O}(\Delta t^2)\nonumber\\
    =&e^{i\Delta t u_{j+1}\hat{B}}\comm{\hat{B}}
                {\hat{C}}e^{-i\Delta t u_{j+1}\hat{B}}+\mathcal{O}(\Delta t^2).
\end{align}
This makes $\tilde{\phi}$ an easy to prepare and measure quantity
\begin{align}
    \label{eq:tphi_approx}
    &\tilde{\phi}(t_j) = \fder{J(t_{j+1})}{u(t_j)}
    \\\nonumber&
        = i\Delta t\bra{\psi_j}
            e^{i\Delta t u_{j+1}\hat{B}}\comm{\hat{B}}
                {\hat{C}}e^{-i\Delta t u_{j+1}\hat{B}}
        \ket{\psi_j}+\mathcal{O}(\Delta t^3).
\end{align}

This quantity, Eq.~(\ref{eq:tphi_approx}), is not directly related to $\phi(t_j)$ anymore.  In specific settings, it might be possible to estimate $\tilde{\phi}$ via $\phi$ and $u$, but it would only lead to one more measurement for each layer of the ansatz to actually just measure $\tilde{\phi}(t_j)$ directly.  Furthermore, most computers capable of measuring $\phi_j$ should also be able to prepare and measure $\tilde{\phi}_j$.

Using Eqs.~(\ref{eq:Phi_approx})~\&~(\ref{eq:tphi_approx}), we can perturbatively estimate the optimal control gradients and therefore find information about how much we should update previous steps in our ansatz based off later steps.

As we said, this currently assumes that all the $\phi$, $\tilde{\phi}$, and $u$ are exactly as they were when we originally measured or prepared them.  If we go about updating previous layers, these formulae will become worse and worse.  Additionally, since this is all perturbative, we cannot apply these approximations for large $(t_f-t_j)$.  For both these reasons, it is sensible to only apply these recursive updates in a short time window behind the current time.  In the next section we will further discuss the breakdown of these approximations and how to deal with them in practice.

As a final note regarding the sampling requirements of these methods on real quantum hardware, the original FALQON required sampling sufficient to determine the sign and rough magnitude of the $\phi$.  Methodologies based off the perturbative approach embodied by Eq.~(\ref{eq:Phi_approx}) do have a stricter sampling requirement in that we now must take enough samples to identify both the sign of the $\phi$ and also the sign of the difference between $\phi$ and $\tilde{\phi}$.  While this does require more samples, these measurements do not need to be horribly exact to carry out the procedure.  Future work will examine the exact sampling requirements for such feedback based algorithms in a quantum setting.

\section{Algorithms}
\label{sec:algorithms}

We now give concrete realizations of our algorithm. As evident from the analysis above, our approach can be applied to 
the quantum alternating operator ansatz for a wide variety of problems, subject to the mild requirement that we can efficiently measure and estimate the required operator commutator expectation values. 
In terms of our derivation, two important considerations for 
our approximations 
discussed above 
are 
\begin{enumerate}
    \item This is perturbative, so we are less certain about steps $j$ for which $(t_f-t_j)^2$ is not small.
    \item As we update the parameters, our estimates for $\phi$ and $\tilde{\phi}$ become less accurate because the control functions leading to those parameters have perturbatively changed.
\end{enumerate}
While both of these aspects are fairly serious limitations, there are multiple ways of dealing with one or both.

One of the obvious solutions to the first point is to only update parameters in some window behind the current time.  This solves the first point nicely and also somewhat addresses the second since changes to very early parameters will not happen, avoiding messing up later steps.

Another solution, addressing the second point is to periodically remeasure $\phi$ and $\tilde{\phi}$ with the new $u_j$ values.  This obviously will add more measurements to the algorithm, but there is nothing fundamentally difficult about these kind of measurements or circuit preparations.

\subsection{Core Algorithms}

Therefore, we have two main algorithmic proposals for addressing the limitations, each of which is based on this notion of a time window
\begin{enumerate}
    \item \{Falling-off\} We keep a time window in which we update the control function, so that we are only using gradient descent to update the previous $W$ layers.  Every time we add a new layer, we also do one step of gradient descent, updating the previous $W$ time layers using approximations of $\Phi_j$ updated with new measurements.  This gradient descent step could be relatively large since we are doing relatively few gradient descent steps.  We also will have the strength of the gradient descent fall off the farther back we go in time, and we discuss the nature of this fall-off in future sections.  This helps both because we are less certain about older steps and so that modifications to those older steps do not unduly change the accuracy of taken measurements on newer steps.
    \item \{Windowed\} Keep a fixed time window in which we do multiple gradient descent steps for all elements in the window, making new measurements of $\phi$ and $\tilde{\phi}$ at every step of gradient descent.  Only after the gradient descent has sufficiently converged, with new measurements of the gradient being close to zero, do we move on to adding the next layer.  For this, it might be sensible to hold off applying gradient descent steps until after a few layers have been added.  This does require more measurements, but it would mean that the only uncertainty is that caused by perturbation theory.
\end{enumerate}
In practice, we have had more success with the Falling-off version of this algorithms; although, it does require fine tuning of the falling-off parameters.  The windowed version's additional measurements also make it costlier but only in a limited way.  In the numerics section, we focus exclusively on the Falling-off version since it performs better and requires fewer resources.  We term these algorithms Feedback Optimally Controlled Quantum States (FOCQS).

\subsection{Continuity of Controls}

One unexpected, but understandable, outcome of our numeric experiments is that these perturbative approaches work better with control functions that are roughly continuous.  Bang-bang approaches are characterized by wide jumps in derivatives, so it is understandable in hindsight that such bang-bang approaches would be poor choices for improvement with a perturbative scheme.

One result of this is that FOCQS-based schemes perform fairly poorly on bounded versions of FALQON (see App.~\ref{app:bounded}) which tend to favor bang-bang regimes more readily.  The drastic jumps seen here mess with the perturbative method and can lead to erratic results, sometimes performing worse than base bounded FALQON.  We provided these bounded versions since they might be interesting in their own right and often perform well, but for our perturbative methods, an unbounded version or methodology that produces a smooth control parameter are preferred.  Due to this erratic nature, the numerics are not horribly edifying.

The original unbounded formulation of FALQON does create continuous control schedules, and as we show in the numerics section, is quite amenable to perturbative enhancement via FOCQS.

The key takeaway here is that these methods are much better dealing with continuous control fields, and care should be taken when applying them.  Note that just because the control function is continuous does not mean the circuit needs to be.  Trotterization of the circuit is still allowable and fine so long as the control function itself can be described in a continuous manner.

\subsection{Falling-Off FOCQS}

\begin{algorithm}
\caption{Falling-Off FOCQS\\(as implemented)}
\label{alg:focqs}
\SetKwInOut{Input}{input}
\Input{$\hat{C}$ - Problem Hamiltonian}
\Input{$\hat{B}$ - Mixer Hamiltonian}
\Input{$T$ - depth of algorithm}
\Input{$\Delta t$ - time step}
\Input{$\beta_0$ - perturbation strength}
$u_0\gets0$\;
\For{$i\gets 0\ldots T$}{
    \For{$j\gets 0\ldots i$}{
        $\hat{H}_j\gets u_j \hat{B} + \hat{C}$\;
    }
    Prepare $\ket{\psi_i} \gets \left[\prod_{k=0}^i e^{-i \Delta t \hat{H}_k}\right]\ket{\varphi_0}$ (Eq.~\ref{eq:state})\;
    Measure $\phi_i\gets i\Delta t\bra{\psi_i}\comm{\hat{C}}{\hat{B}}\ket{\psi_i}$ (Eq.~\ref{eq:phi_j0})\;
    $u_{i+1}\gets\max(0, -\phi_i)$ (Eq.~\ref{eq:FALQON_con_0})\;
    Prepare $\ket{\tilde{\psi}_i} \gets e^{-i\Delta t u_{i+1}\hat{B}}\ket{\psi_i}$\;
    Measure $\tilde{\phi}_i \gets i\Delta t\bra{\tilde{\psi}_i}\comm{\hat{B}}{\hat{C}}\ket{\tilde{\psi}_i}$ (Eq.~\ref{eq:tphi_approx})\;
    \For{$j\gets 0\ldots i-1$}{
        $\Phi_j \gets \phi_j+(i-j)(\tilde{\phi}_j-\phi_j)$ (Eq.~\ref{eq:Phi_approx})\;
        $u_j\gets u_j + \Phi_j/(\beta_0(i-j)^2)$ (Eq.~\ref{eq:focqs_u_update})\;
    }
}
\end{algorithm}

We next consider implementation of our algorithms, directly
comparing the results to the original formulation of FALQON.  Surprisingly, we get better results with Falling-Off as described above than with Windowed.  There are a number of meta-parameters and additional details to be described, so before we get into the numeric results, we describe the details of the Falling-Off FOCQS algorithm.  The pseudocode for this implementation of the algorithm is displayed in Alg.~\ref{alg:focqs}.

FOCQS operates first and foremost like FALQON as described in Section \ref{sec:FALQON}.

The innovation in Falling-Off FOCQS is that we update the previous $u_j$ whenever we add a new layer.  To do this, we estimate $\Phi_j$ for each layer using Eq.~\ref{eq:Phi_approx}.  We then update the $u_j$ so that (denoting the most recent layer with $k$)
\begin{equation}
    \label{eq:focqs_u_update}
    u_j \to \begin{cases}
        u_j & (k-j)>W\\
        u_j + \Phi_j/(\beta_0(k-j)^f) & \text{o/w}
    \end{cases}
\end{equation}

We have three new meta-parameters here, $W$, $\beta_0$, and $f$.  $W$ is the size of the window that we are updating in, $f$ is the fall-off coefficient, and $\beta_0$ gauges the base step size of the gradient descent.

If $f$ is chosen appropriately, then layers too far in the past won't be updated much at all, so with an appropriate $f$, we find that $W$ can be set to be as large as the entire procedure with no ill-effects.  Furthermore, the size of the window doesn't matter for quantum computational time since we already have all the measurements needed to calculate $\Phi_j$ at each step.  This does add some extra classical complexity dependence on the number of FOCQS layers, but it is only a linear dependence on the circuit depth and can be done efficiently with classical resources.

Perhaps unsurprisingly, our testing indicates that $f=2$ is a good choice for the fall-off coefficient.  This is unsurprising to us because our error terms in the calculations of $\Phi_j$ are quadratic in the quantity $(k-j)$, so we become less certain in our estimates going back in time in a quadratic manner.  $f>2$ certainly still works but tends to be too conservative in our simulations, producing results that aren't too far away from FALQON.  Although, even $f\to\infty$, meaning updating just one step back, provides a noticeable improvement over FALQON.  $f<2$ tends to lead to situations where the measured and actual $\phi_j$ and $\tilde{\phi}_j$ differ from each other too much, leading to protocols that become pathological as the algorithm breaks down.  This break down can be corrected with a finite $W$, but in general ignoring $W$ and setting $f\approx 2$ performs better than artificially restricting the window.

$\beta_0$ is then the main meta-parameter and should be modified based on system size and problem type.  In the results below, we report our values of $\beta_0$ that were determined through meta-parameter tuning by hand.  A higher $\beta_0$ corresponds to a smaller gradient descent step.  Because we are running gradient descent for a finite number of steps, not until convergence, it is likely that smaller $\beta_0$ are acceptable in this setting.  In general, we found that $\beta_0$ in the range roughly 10 to 20 worked well.  $\beta_0$ lower than 10 can sometimes work very well, but it can sometimes lead to procedures that lose track of their goal entirely.  Higher $\beta_0$ work and are stable, but they just don't lead to as much advantage.

\subsection{Iterative Procedure}

Notice additionally that the update procedures described just rely on control theory for updating parameters, not on control theory for the original selection of parameters.  FALQON, of course, based its parameter selection rules on Lyapunov control theory, but this is a distinct procedure from how gradient measurements are used to perturbatively update past parameters.

In fact, these two notions of initial parameters selection and perturbative parameter updating can be decoupled from each other.  A simple example of this would be to consider a fixed discretized annealing schedule, such as would often be found in a discretized adiabatic algorithm.  Before going into the procedure, there would be some initial set of $u^{(0)}_i$ parameters selected.  This iterative perturbative procedure would then proceed through building up the circuit one layer at a time, making measurements of $\phi(t_i)$ and $\tilde{\phi}(t_i)$ at the end of a length $i$ circuit.  The next value of the new control parameter, $u^{(1)}_i$ would then be set to $u^{(0)}_i$ rather than using FALQON's prescription.  But then our perturbative approach would still be used to update the values of $u^{(1)}$ as described in the previous subsection.  This will produce a circuit that is perturbatively different from the one defined by $u^{(0)}_i$ alone.  Whether or not this leads to a major improvement depends intimately on the initial choice of $u^{(0)}$ as well as the structure of the control landscape.  Pseudocode for this iterative procedure is provided in Alg.~\ref{alg:iterative}.

Of course, since this procedure is decoupled from the parameter selection itself, this can be used in an iterative fashion, using $u^{(1)}$ itself to generate $u^{(2)}$ and so on.  We develop this methodology further in our numerics section, using a basic version of FOCQS to generate $u^{(0)}$ and then iteratively updating these parameters to refine the procedure.  In practice, we find that this iterative procedure leads to steady but diminishing improvements.

\begin{algorithm}
\caption{Iterative FOCQS\\(as implemented)}
\label{alg:iterative}
\SetKwInOut{Input}{input}
\Input{$\hat{C}$ - Problem Hamiltonian}
\Input{$\hat{B}$ - Mixer Hamiltonian}
\Input{$T$ - depth of algorithm}
\Input{$\Delta t$ - time step}
\Input{$\beta_0$ - perturbation strength}
\Input{Existing smooth procedure, $u_{j}^{(0)}$ for $j \in[0,T]$}
\For{$i\gets 0\ldots T$}{
    $u_i\gets u_i^{(0)}$\;
    \For{$j\gets 0\ldots i$}{
        $\hat{H}_j\gets u_j \hat{B} + \hat{C}$\;
    }
    Prepare $\ket{\psi_i} \gets \left[\prod_{k=0}^i e^{-i \Delta t \hat{H}_k}\right]\ket{\varphi_0}$ (Eq.~\ref{eq:state})\;
    Measure $\phi_i\gets i\Delta t\bra{\psi_i}\comm{\hat{C}}{\hat{B}}\ket{\psi_i}$ (Eq.~\ref{eq:phi_j0})\;
    Prepare $\ket{\tilde{\psi}_i} \gets e^{-i\Delta t u^{(0)}_{i+1}\hat{B}}\ket{\psi_i}$\;
    Measure $\tilde{\phi}_i \gets i\Delta t\bra{\tilde{\psi}_i}\comm{\hat{B}}{\hat{C}}\ket{\tilde{\psi}_i}$ (Eq.~\ref{eq:tphi_approx})\;
    \For{$j\gets 0\ldots i-1$}{
        $\Phi_j \gets \phi_j+(i-j)(\tilde{\phi}_j-\phi_j)$ (Eq.~\ref{eq:Phi_approx})\;
        $u_j\gets u_j + \Phi_j/(\beta_0(i-j)^2)$ (Eq.~\ref{eq:focqs_u_update})\;
    }
}
\end{algorithm}

\subsection{Extensions to constrained optimization}

Here we explicitly remark on the application of 
feedback-based quantum optimization to problems with hard constraints. Recall that our analysis and presentation has been mostly general so far, with the Hamiltonians $\hat{C}$ and $\hat{B}$ left unspecified. 
In a \emph{constrained} optimization problem, we again seek to minimize a classical cost Hamiltonain $\hat{C}$, but now subject to an additional set of constraints being satisfied, which can without loss of generality be expressed as $\hat{A}\ket{\psi}=0$. Generally speaking there exist two distinct approaches to modifying existing quantum algorithms such as QAOA for this setting which we now describe. 

In the first approach~\cite{hen2016quantum,Hadfield2019}, the mixing Hamiltonian $\hat{B}$ and initial quantum state are modified in a problem-dependent way so as to ensure that the quantum state produced only ever contains support on computational basis states corresponding to feasible problem solutions. In this case, it is easy to see that both FALQON and our FOCQS extension can be applied directly with minimal adaptation. Here, quantities such as $\comm{\hat{C}}{\hat{B}}$ will change appropriately according the particular problem and mixer at hand. 

For the second approach, closely related to methods from quantum annealing, the mixer $\hat{B}$ is retained as the transverse field, and instead the cost Hamiltonian $\hat{C}$ is augmented with additional terms that provide an energy penalty to any infeasible states. While this setup is also quite general, care must be taken as to how penalty terms and weights are designed, especially in the setting of approximate optimization. 

While both approaches are applicable to a wide variety of problems~\cite{lucas2014ising,Hadfield2019}, in our numerical experiments below we apply only the second option as it is relatively generic, and facilitates direct comparison to unconstrained problems tackled with the same quantum circuit (mixing operator). There we consider an important constrained optimization problem, Maximum Independent Set. In future work we will perform a detailed investigation of tradeoffs for additional classes of hard constrained optimization problems and different quantum circuit ans\"atze.

\section{Numerics}
\label{sec:numerics}

For numerics, we focus on noiseless quantum simulations using classical computers which limits our system sizes but allows us to focus on the algorithms and not directly on noise.  Our classical simulator uses direct state evolution via a Trotterization of the continuous control problem.

Below we also describe the models that we run numerics on, using both the Falling-Off FOCQS described in Alg.~\ref{alg:focqs} and the iterative procedure described in Alg.~\ref{alg:iterative} applied to the output control function of FOCQS.

\subsection{Models}

We seek to test both FOCQS and its iterative version using classical simulations.  To do this, we will employ two different examples, trying to capture different regimes in which FALQON might serve to be improved.  Both of these problems are quadratic binary optimization problems due to the ease of implementing these on near-term quantum hardware, but we include a model that is constrained (implemented through penalty terms), where base FALQON potentially struggles but where other modifications of FALQON have been proposed \cite{Rahman2024}.

In both these models, we use a transverse field on $n$ qubits as the mixer Hamiltonian:
\begin{equation}
    \hat{B} = -\sum_{i=1}^n \sigma_x^{(i)}.
\end{equation}
We use the notation $\sigma_\alpha^{(i)}$ to indicate the $\alpha$ Pauli matrix acting on qubit $i$.

\subsubsection{Ising Spin-Glass/MaxCut}

The Ising spin-glass is equivalent to a graph MaxCut problem with weighted edges, generally referred to as a Quadratic Unconstrained Binary Optimization (QUBO) problem.  The Hamiltonian can be written as
\begin{equation} \label{eqn:maxcut}
    \hat{C} = \sum_{i=1}^n\sum_{j=i+1}^n J_{ij}\sigma_z^{(i)}\sigma_z^{(j)}.
\end{equation}
We choose an all-to-all connected graph with each $J_{ij}$ chosen uniformly at random from the range $[-1,1]$.

As a graph problem, finding the ground state of this Hamiltonian represents trying to sort the nodes of the graph into two groups such that as many edges between the two groups are cut as possible (since this is weighted with some negative edge weights, we want those negative edges to not be cut).

We emphasize that for QUBO problems, possibly including linear terms, and the transverse-field mixer, the commutator $[\hat{B},\hat{C}]$ is simple to compute and contains at most twice the number of terms in the Pauli basis as~$\hat{C}$. 
In particular, for Eq.~\eqref{eqn:maxcut} we have 
$[\hat{B},\hat{C}] = 2i\sum_{i=1}^n\sum_{j=i+1}^n J_{ij}(\sigma_y^{(i)}\sigma_z^{(j)}+\sigma_z^{(i)}\sigma_y^{(j)})$.
While the resulting terms do not in general mutually commute, they can be measured in a straightforward way for example utilizing a term partitioning scheme~\cite{tilly2022variational,kurita2023pauli}.

\subsubsection{Maximum Independent Set}

Maximum Independent Set is again a graph problem on an $n$ node graph.  Here the goal is to find the maximum size set of nodes such that no nodes in the set have an edge between them.  We consider specifically the weighted version of this problem, where node is given a weight, with the goal to find the independent set with the maximum sum of weights of nodes in that set.  For our implementation, we take these weights, $r_i$, to be randomly distributed in the range $[0,2]$.  Using a penalty formulation, the Hamiltonian can be written in two pieces: a reward, $\hat{C}_r$ and a penalty $\hat{C}_p$.  Furthermore, a penalty weight $\lambda$ (chosen to be $2$ in our numerics) is used such that:
\begin{equation}
    \hat{C} = \hat{C}_r + \lambda \hat{C}_p,
\end{equation}
where the goal is to find a configuration that minimizes $\hat{C}_r$ while being in the ground space of $\hat{C}_p$.  We want to formulate this as a minimization problem, so we include the rewards as a negative.  This reward Hamiltonian is just a simple reward for each node in the set:
\begin{equation}
    \hat{C}_r = -\sum_{i=1}^n \frac{r_i}{2} (\sigma_z^{(i)}+I),
\end{equation}
which is scaled to give the reward if the qubit is in the $+1$ state and nothing if it is in the $-1$ state.

For the penalty term, we want the Hamiltonian to be zero if there is no edge between qubits set to $+1$, and we want a positive penalty if there is such an edge:
\begin{equation}
    \hat{C}_p = \sum_{i=1}^n \sum_{j=i+1}^n \frac{1}{4}(\sigma_z^{(i)}+I)(\sigma_z^{(j)}+I),
\end{equation}
where we have scaled each penalized edge to have weight one.

\subsection{Results}

\begin{figure}
    \begin{center}
    \includegraphics[width = 0.475\textwidth]{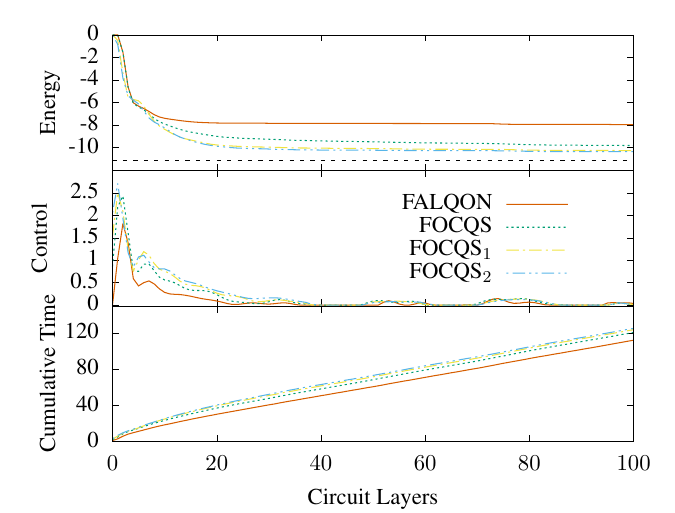}
    \end{center}
    \caption{Simulation details for a random Ising spin-glass problem on $n=10$ qubits.  All plots show the progression of the model as a function of circuit layers, considering the final version of the control, after the FOCQS procedure is complete.  The top panel shows the achieved energy with the dashed black line showing the true ground state energy of this problem.  The middle panel shows the control function $u_j$ settled on by the completed procedures.  The bottom panel shows the cumulative time that the procedure takes, assuming that each problem gate takes one time unit and each mixer gate takes a time equal to its control parameter.  In this example $\beta_0=10$ was taken, and additional iterative versions of FOCQS are considered.
    }
    \label{fig:stacked_example}
\end{figure}

First across all these examples, our actual results tend to be qualitatively similar, so we have plotted an exemplar run in Fig.~\ref{fig:stacked_example}.  This was a random Ising instance on $n=10$ qubits run for 100 layers of FALQON and FOCQS along with two iterated versions of FOCQS denoted by FOCQS$_i$ with FOCQS$_1$ using as input the control function determined by FOCQS and FOCQS$_2$ using FOCQS$_1$ as input.  In this plot $\beta_0=10$ and $\Delta t = 0.1$.  This simulation and all other simulations were carried out with noiseless state evolution using a custom designed and optimized simulator.

The three panels in Fig.~\ref{fig:stacked_example} each depict different aspects of the same simulations.  The top panel shows the instantaneous energy achieved by the final versions of the procedures with the dashed black line representing the true ground state energy.  Notice first that all the procedures get stuck in a local minimum but that FOCQS and its iterative versions manage to get to much lower energies than base FALQON even if they do not seemingly reach this result faster than FALQON.  Qualitatively this behavior was seen across all our numerics with FOCQS and FALQON tracking each other roughly at the beginning before diverging as various procedures get stuck in minima.  In addition, as can be seen from this plot, 100 circuit layers is more than enough for a seemingly stable minima to be found in this and all other examples.

The second panel shows the value of the control $u_j$ during these iterations.  $u_0=0$ is mandated at the beginning because any application of the mixer is useless when starting in an eigenstate of the mixer \cite{Brady2021}.  But right after this, strong application of the mixer seems preferred with each subsequent iteration increasing this initial jump.  The nature of these curves does bear some resemblence to the optimal bang-anneal-bang curves that result from Pontryagin optimal control \cite{Brady2021,Brady2021} with this mandated bang at the beginning, followed by a jump up to the mixer and an anneal down to the problem, but without the final bang.  Obviously, our methods are just a perturbative feedback manner of trying to approximate those optimal control methods, so we cannot expect exact convergence to those optimal procedures.

The final panel just shows the rough cumulative time the procedures take.  One argument that could be leveled against FOCQS and its iterative forms is that the runtime (or alternatively energy) required for the procedure is being pushed upward since the control parameter correlates to time (energy) usage.  Therefore, seemingly FOCQS is achieving better results with more time, and this panel is designed to display how much more time FOCQS is using.  This time is calculated assuming that it take FALQON and FOCQS a single time unit to apply the problem Hamiltonian at each layer.  Further, the time required to apply the mixer Hamiltonian is taken to be equal to the control function value, $u_j$, quantifying how many applications of gates corresponding to that mixer would be needed.  
In the end, the result is that all of the algorithms are taking roughly equivalent amounts of time (alternatively, energy), so the increased performance of FOCQS and its iterative variants is not solely due to increased runtime but instead better usage of that time.

In the next few plots we show averaged results for our different models.  For each $n$ value listed, we generated 50 random problems as described above.  The data points show the approximation ratio, defined here as the achieved energy by the end of the procedure ($100$ circuit layers and $\Delta t = 0.1$) divided by the true ground state energy.  
For both problems a value of one represents solving the problem exactly with lower values representing sub-optimal solutions.
The error bars show the standard error of the mean across our simulations.

\begin{figure}
    \begin{center}
    \includegraphics[width = 0.475\textwidth]{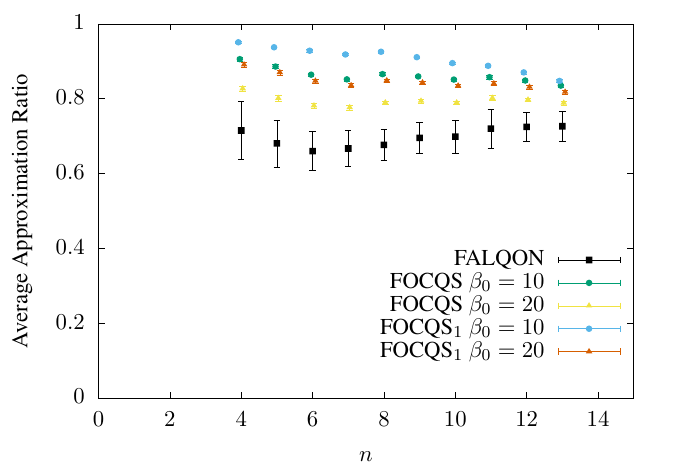}
    \end{center}
    \label{fig:MC_app_ratio}
    \caption{Across 50 trials for each $n$ value of randomly chosen Ising spin-glasses, we calculated the best approximation ratio (achieved energy divided by true ground state energy) achieved in the procedure.  This plot shows the average of those approximation ratios versus $n$ with the error bars representing the standard error of the mean of these approximation ratios.  We show two different values of $\beta_0$, and the subscript on FOCQS$_1$ refers to the iterative version of FOCQS applied for one iteration past normal FOCQS.}
\end{figure}

Fig.~\ref{fig:MC_app_ratio} shows the averaged results for the Ising problem example.  Note first that each iteration of FOCQS improves on FALQON or previous iterations, which is expected of perturbative methods.  For $\beta_0$, the more conservative $\beta_0 = 20$ still gives an improvement but to a lesser degree than $\beta_0=10$.  These $\beta_0$ values are quite small for gradient descent, corresponding to fairly large gradient steps.  This largness of step size can be explained by the fact that we are not running gradient descent to convergence but just for a few steps, where we are assumed and in practice far from the optimum, meaning overshooting the optimum is not a concern.  One of the more interesting results of FOCQS is that it severely decreases the spread of simulation results as shown in the decreased error bars. 
This decrease in the spread indicates that FOCQS can take advantage of easy optimization of FALQON, bringing all results closer together; although, a more formal explanation is still needed.

\begin{figure}
    \begin{center}
    \includegraphics[width = 0.475\textwidth]{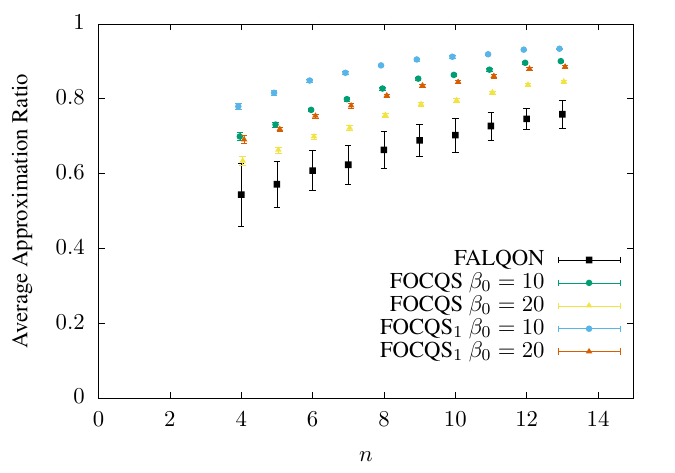}
    \end{center}
    \label{fig:MIS_app_ratio}
    \caption{Across 50 trials of random weighted Maximum Independent Set problems on Erd\"os-R\'enyi graphs of size $n$, we calculated the best approximation ratio (achieved energy divided by true ground state energy) achieved in the procedure.  This plot shows the average of those approximation ratios versus $n$ with the error bars representing the standard error of the mean of these approximation ratios.  We show two different values of $\beta_0$, and the subscript on FOCQS$_1$ refers to the iterative version of FOCQS applied for one iteration past normal FOCQS.}
\end{figure}

For Maximum Independent Set, Fig.~\ref{fig:MIS_app_ratio}, we use random, Erd\"os-R\'enyi graphs with the probability of any edge existing being $1.2\ln(n)/n$, and we further accept only graphs that do not have disconnected subgraphs.  Much of the results here are the same as in the previous example.  Note that the increasing approximation ratio with $n$ is not expected to hold indefinitely and is likely a small size side effect.

Overall, our results show an improvement of FOCQS and its iterated versions over base FALQON.  These methods are perturbative, so this improvement is not unexpected.  Unfortunately, their perturbative nature also limits their effectiveness with the true ground state still out of reach.

\section{Conclusions}
\label{sec:conclusion}

FALQON and related feedback-based approaches represent an important advancement for quantum computing as one of the first attempts to create non-variational schedules for optimization algorithms that can still recover some of the advantage of variational quantum algorithms while avoiding the significant overhead of parameter optimization. 
On the other hand, as explained feedback-based algorithms such as FALQON come with various caveats, not least their lack of guarantees that they can reach the true ground state; although, similar guarantees are also generally lacking for variational quantum algorithms in practice with specific initial guesses, limited circiuit depth, and bounded classical optimization loops.  This work provides a way of improving on feedback-based approaches by introducing FOCQS, a perturbative scheme for improving on the results of FALQON without significantly more measurements.  Our FOCQS methods also provides a way of integrating Pontryagin and Lyapunov control theory in a way that is not sensible in classical settings but is reasonable in the quantum setting where each measurement necessitates reconstructing the state for the next measurement.

Naturally there are also caveats to our algorithm.  Due to its perturbative nature, its improvements are incremental, but this can be mitigated by iterative applications of the algorithm.  In fact, the general FOCQS structure can be applied iteratively to any annealing-like control schedule independent of whether it came from FALQON or not.  Our results do indicate that this method breaks down for control functions with many discontinuities, and we leave a detailed study of control functions to future research.

There is still much room for improvement in these non-variational quantum annealing and QAOA like algorithms, as well as extensions of these techniques to wider variety of cost Hamiltonians, including so-called quantum problems (i.e., non-diagonal cost Hamiltonians)~\cite{ho2019efficient,kremenetski2021quantum}.
Indeed, several recent work have considered application of FALQON to molecular Hamiltonians from quantum chemistry as well as the Fermi-Hubbard model~\cite{larsen2023feedback,rahman2024feedback,pexe2024using}.
Our control theory approach provides a general method for potentially improving these algorithms and a framework in which to continue such feedback-based extensions. 

Finally, we remark on an important closely related class of protocols, so called adaptive methods such as ADAPT-VQE~\cite{grimsley2019adaptive} and ADAPT-QAOA~\cite{zhu2022adaptive,yanakiev2024dynamic}. In these protocols, ans\"atze are again constructed layer by layer using data drawn from the quantum circuit, with a key difference being the pool of potential (mixing) operators to add at each layer now contains more than one choice. As this phrasing suggests, the control theory perspective of this work as well as tools from small-angle analysis~\cite{hadfield2022analytical} seem well-suited to obtain similar improvements in the adaptive setting. More generally, we are optimistic that these tools will lead to further inroads into the many related open questions regarding performance and the complexity of parameter setting for variational quantum circuits.

\acknowledgments
We thank Zoe Gonzalez Izquierdo, Davide Venturelli, P. Aaron Lott, and Eleanor Rieffel for helpful discussions and/or guidance on this paper. 
We are grateful for support from NASA Ames Research Center. 
This work was supported by the Defense Advanced Research Projects Agency (DARPA) under Agreement No. HR00112090058 and DARPA-NASA IAA 8839, Annex 114.  S.H. was supported by NASA Academic Mission Services under contract No. NNA16BD14C.

\bibliography{refs}

\begin{appendix}

\section{Bounded FALQON and FOCQS}
\label{app:bounded}

\subsection{Bounded FALQON}
Here we present a minor modification of FALQON to make it more in line with the standard notations of QAOA and Quantum Annealing while simultaneously imposing realistic constraints on the strengths of the control fields applied.

For this bounded FALQON, we can go to a more standard QAOA or quantum annealing style Hamiltonian that looks like
\begin{equation}
    \hat{H}_j = u_j \hat{B} + (1-u_j) \hat{C}
\end{equation}
where $u_j\in[0,1]$. 
Officially nothing changes in the derivative in Eq.~(\ref{eq:phi_j}) from before, but it will be convenient later on (cf. Eq. (\ref{eq:boundedPhi})) to explicitly include an additional 
factor of $\hat{C}$

\begin{equation}
    \label{eq:phi_j_2}
    \phi_j = \phi(t_j) = i\Delta t\bra{\varphi_0}\comm{\hat{C}(t_j)}{\hat{B}(t_j)-\hat{C}(t_j)}\ket{\varphi_0}.
\end{equation}

We will still set the control function based off $\phi_j$, but in a slightly different way, mostly due to the fact that our $u(t) \in [0,1]$ is bounded. 

Given the bounded nature of our results, we will consider two different possible formulations, the first of which is most similar to the original FALQON,
\begin{equation}
    \label{eq:FALQON_con_1}
    u_{j+1} = 
        \begin{cases}
            0 & \phi_j>0\\
            -\phi_j & -1<\phi_j<0\\
            1 & \phi_j<-1
        \end{cases},
\end{equation}
and the second of which is more bang-bang inspired,
\begin{equation}
    \label{eq:FALQON_con_2}
    u_{j+1} = 
        \begin{cases}
            0 & \phi_j>0\\
            1 & \phi_j\leq0
        \end{cases}.
\end{equation}

We present both of these for completeness, but in practice the version with non-bang-bang controls performs better and is easier to handle perturbatively as we seek to do later.

Furthermore, this bounded version of FALQON often leads to sharp discontinuities in the control parameter.  These discontinuities are not a problem for FALQON itself and often produce good-quality procedures, but our perturbative method struggles with such control discontinuities meaning that most of our perturbative methods will be numerically tested with the original unbounded form of FALQON.  Still to our knowledge, this bounded form of FALQON has never been explicitly expressed in literature, so we provide it for further potential study.

\subsection{Optimal Control}

Here we discuss the modifications to optimal control functions and FOCQS that would be necessary to account for a bounded control function.

The functional derivative of the objective function with respect to the control function (cf. Eq.~(\ref{eq:Phij}) becomes
\begin{equation}
    \label{eq:boundedPhi}
    \Phi_j(t_f) = \Phi(t_j,t_f) = i\Delta t\bra{\varphi_0}\comm{\hat{C}(t_f)}{\hat{B}(t_j)-\hat{C}(t_j)}\ket{\varphi_0}.
\end{equation}

The majority of the rest of our derivations apply, just using bounded versions of $\phi_j$ and $\Phi_j$.  The only main difference is that $\tilde{\phi}_j$ needs to be rederived in the bounded setting.  The results are similar to the unbounded setting, Eq.~\ref{eq:tphi_approx}, just with an extra $\hat{C}$ floating around:
\begin{align}
    \label{eq:tphi_approx_bounded}
    &\tilde{\phi}(t_j) = \fder{J(t_{j+1})}{u(t_j)}
    \\\nonumber&
        = i\Delta t\bra{\psi_j}
            e^{i\Delta t u_{j+1}(\hat{B}-\hat{C})}\comm{\hat{B}}
                {\hat{C}}e^{-i\Delta t u_{j+1}(\hat{B}-\hat{C})}
        \ket{\psi_j}
        \\\nonumber
        &+\mathcal{O}(\Delta t^3).
\end{align}
This measurement does involve an evolution using time evolution generated via $\hat{B}-\hat{C}$ which is not assumed to be possible in the bounded setting.  Such evolution should be possible in most physical systems with bounded controls but will require Trotterization schemes.

With modifications to just these quantities, we could apply FOCQS to a bounded feedback algorithm.  As we discuss in the main text, FOCQS works better in a setting with roughly continuous control functions, which the bounded version of FALQON often does not provide.  Therefore, while the core methodology of FOCQS works with bounded control functions in theory, we do not recommend this in practice.

\end{appendix}

\end{document}